Defect physics of BaCuChF (Ch = S, Se, Te) *p*-type transparent conductors

A. Zakutayev, J. Tate, and G. Schneider[†]

Department of Physics, Oregon State University, Corvallis, OR 97331, USA

Abstract

Native point defects, defect complexes, and oxygen impurities in BaCuChF were studied using density functional theory calculations, self-consistent thermodynamic simulations, and various experimental techniques. Unintentional *p*-type conductivity in BaCuChF is explained by the presence of copper vacancies with transition levels in the valence band. These acceptor-like defects are partially compensated by donor-like chalcogen vacancies with transition levels deep in the gap. Chalcogen vacancies also cause the experimentally observed sub-gap photoluminescence, optical absorption, and persistent photoconductivity in BaCuSF and BaCuSeF. In thermodynamic equilibrium, both copper and chalcogen vacancies have low formation enthalpies and are likely to form defect complexes among themselves and with fluorine interstitials. The calculated Fermi level pinning range in BaCuChF is narrow and located close to the valence band maximum. It makes BaCuChF a suitable transparent *p*-type contact layer for optoelectronic applications, but hinders attempts to fabricate transparent thin film transistors using this material. Oxygen-related defects do not affect bulk BaCuChF properties, but surface oxidation decreases the mean free path of free holes by almost an order of magnitude.

I. INTRODUCTION

BaCuChF (Ch = S, Se, Te) is a family of *p*-type transparent conductors[1] suitable for contact applications in thin film solar cells[2] and organic light-emitting devices.[3] These fluorochalcogenides are isostructural[4,5,6] with the oxychalcogenide *p*-type transparent conductors LnCuOCh (Ln = La, Pr, Nd)[7] and the oxypnictide high-temperature superconductors LaFeOPn (Pn = P, As, Sb)[8] (Fig. 1). Understanding the origin of *p*-type conductivity in BaCuChF and LaCuOCh is important for further progress in the search for a good *p*-type transparent conductor. In 40-nm LaCuOSe thin films, *p*-type conductivity is attributed to copper vacancies,[9] which was confirmed by recent density functional theory calculations of the formation energy of a neutral Cu vacancy in LaCuOSe.[10] However there are no systematic theoretical studies on the origin of *p*-type conductivity in either BaCuChF or LaCuOCh.

Density functional theory (DFT) provides a theoretical way to examine the physics of defects in materials. DFT calculations have been used in the literature to explain the properties of intrinsic defects and extrinsic impurities in binary (for example CdTe,[11] $In_2O_3$,[12] ZnO[13,14]) and ternary (for example $CuInSe_2$,[15,16] $CuAlO_2$,[17,18] $CuSr_2O_2$[19]) compounds. The defect physics of quaternary semiconductors such as $Cu_2ZnSnS_4$ was reported only recently,[20,21] and it is complex due to possible cation cross-substitution. Mixed-anion quaternary semiconductors, such as BaCuChF and LaCuOCh have an even larger variety of possible defects, because of a potential disorder on both the cation and the anion sub-lattices.

We report a study of native point defects, defect complexes and oxygen impurities in BaCuChF (Ch = S, Se, Te) using both experimental and theoretical techniques. The aim of this work is to explain a number of experimental observations and to elucidate the origin of *p*-type conductivity in BaCuChF. According to our DFT calculation results, copper vacancies are the lowest-energy acceptor-like defects and chalcogen vacancies are the lowest-energy donor-like defects in BaCuChF. Oxygen-related defects do not affect bulk properties, but surface oxidation affects the charge transport. Self-consistent thermodynamic simulation results show that partial compensation of copper and chalcogen vacancies results in a narrow Fermi level pinning range close to the valence band maximum and leads to a high concentration of free holes in nominally undoped BaCuChF materials. Experimentally measured electrical transport properties, sub-gap photoluminescence, optical absorption, and persistent photoconductivity are consistent with this theoretical defect model.

Theoretical and experimental methods are described in Section II. The discussion of the results in Section III is arranged as follows: in subsection (A) we consider the phase stability range of BaCuChF, in (B) we discuss the physics of native point defects, in (C) the resulting Fermi levels, and in (D) the optoelectronic properties. Finally, we address in subsection (E) oxygen-related defects and in (F) various defect complexes. Each of these discussion subsections contains both theoretical and experimental results. The paper ends with a summary and conclusions (Section IV).

## II. METHODS

### A. Theory

In this work we use the $A_B^q$ defect notation, where $A$ is the actual occupant of the site ($V$ for a vacancy), $B$ is the nominal occupant of the site ($I$ for an interstitial site), and $q$ is the charge of the defect. Defect transition levels are written using $A_B^{q/q'}$ notation, where $q$ and $q'$ are the two charge states of the defect.

The formation enthalpy $\Delta H_{D,q}$ of a point defect $D$ in a charge state $q$ is calculated as

$$\Delta H_{D,q}(E_F) = (E_{D,q} - E_H) + \sum_a n_a(\mu_a^0 + \Delta\mu_a) + q(E_V + E_F) \tag{1}$$

where $E_{D,q}$-$E_H$ is the difference in the total free energies of the material with and without the defect, $n_a$ is the number of atoms $a$ removed from the host to create the defect, $\mu_a^0+\Delta\mu_a$ is the absolute chemical potential of atomic species $a$ in the material ($\mu_a^0$ is the chemical potential in the elemental substance), and $E_V+E_F$ is the absolute Fermi level ($E_V$ is the energy of the valence band maximum). Total free energies $E_{D,q}$ and $E_H$ and defect transition levels are calculated using a DFT supercell approach.[22] For the chemical potentials the value of $\mu_a^0$ and range of $\Delta\mu_a$ are calculated from the DFT formation energies of BaCuChF and its possible impurities.[15] The range of the variable $E_F$ is not limited to the band gap. The theoretical equilibrium Fermi levels at synthesis and room temperature, and corresponding equilibrium concentration of defects and free charge carriers are calculated using self-consistent thermodynamic simulations.[22] The energy cost to form a defect complex $\delta H_{DC,q}$ is calculated as a difference between the sum of formation enthalpies of the isolated defects and formation enthalpy of a defect complex.[15]

B. Computations

DFT supercell calculations were performed using the Perdew-Burke-Ernzerhof (PBE)[23] general gradient approximation (GGA) and the projected augmented wave (PAW) method with a 400 eV energy cut-off.[24] DFT/PBE equilibrium BaCuChF crystal lattice parameters ($a$= 4.149, $c/a$= 2.200 for BaCuSF, $a$= 4.271, $c/a$= 2.169 for BaCuSF, and $a$= 4.464, $c/a$= 2.128 for BaCuSF)[25] were fixed, and all the atoms were allowed to relax without any symmetry constraints during the calculation. The charge density was calculated using the tetrahedron method over a $\Gamma$-centered 4x4x4 $k$-point mesh ($k$ = 36 special points). 2x2x1 supercells ($n$ = 32 atoms) were used for the reference calculations. For the defect calculations, an appropriate number of atoms and electrons were added or removed from the cell. Larger calculations with up to $n = 200$ atoms and $k = 216$ k-points were performed on copper and sulfur vacancies to check the convergence of the results with respect to the size of the supercell and with respect to the number of *k*-points.

Results obtained using the supercell defect calculations were corrected for (i) core level misalignment in the reference and defect calculations, (ii) DFT band gap error, and (iii) defect band formation and filling caused by finite supercell sizes.[22] Potentials of the supercells with- and without the defects were aligned using as a reference Ba 5s states, which are located approximately 30 eV below the valence band maximum (VBM). The conduction band minimum (CBM) was shifted up to match the GW and experimental gaps (3.6 eV for BaCuSF, 3.3 eV for BaCuSeF, and 2.2 eV for BaCuTeF).[26] The same correction was applied to the formation enthalpies of donor-like defects that were not fully charged. Alignment of all peaks in the calculated density of states (DOS) with experimental x-ray photoemission spectroscopy (XPS) spectra[3] allows for a correction of the valence band maximum. However, no single shift was found to lead to satisfactory alignment of all DOS peaks and this correction was not performed. The image charge correction[27] was intentionally omitted.

Convergence checks indicate that after all corrections, the maximum error of the 2x2x1 supercell calculations is 0.25 eV with respect to the extrapolated result for infinite cell size. Hence, 0.25 eV represents the maximum error for all quoted formation enthalpies and Fermi levels. Regardless of the theoretical uncertainty, the S-Se-Te trends of the calculated equilibrium Fermi levels, width, and positions of the Fermi level pinning ranges with respect to the VBM are robust with respect to all corrections.

C. Experiments

Photoluminescence (PL) experiments were performed on BaCuSF powders prepared by solid-state reactions in evacuated silica tubes,[5,6] and on *c*-axis oriented BaCuChF thin films prepared by pulsed laser deposition (PLD) on amorphous fused silica (*a*-$SiO_2$) and single-crystalline 001-cut magnesium oxide (MgO) substrates.[25,28] PL was excited using a $N_2$ laser (337 nm, 120 µJ, 20 Hz, 4 ns pulses) and measured using a spectrometer equipped with a CCD detector. The resulting PL spectra were fit using the sum of Lorentzians. Optical absorption measurements were performed on intentionally S-poor or Cu-rich BaCuChF thin films using a custom-built grating spectrometer. To eliminate the effect of the interference of light from the front and the back of the film, both transmittance and reflectance were accounted for in the calculation of the absorption spectra.[29]

The concentration of free holes was determined from resistivity and Hall effect measurements on patterned BaCuChF thin films in Van der Pauw configuration.[25] From these data we calculated experimental Fermi levels using the DFT density of states and room-temperature Fermi-Dirac distribution.[26] Fermi levels were also directly measured using XPS on polished sintered BaCuChF pressed pellets using monochromatic Al *K*α radiation.[3] XPS is a surface-sensitive technique and measures a surface Fermi level position, in contrast to the bulk Fermi level deduced from the experimentally determined concentration of free holes.

The Seebeck coefficient was measured with a 3-5 °C sample temperature gradient using a custom-built setup. Chemical composition of the samples was measured using electron probe microanalysis (EPMA), and the stoichiometries reported in section III (C) are the results of these studies.

III. RESULTS AND DISCUSSION

A. Phase stability

First, we consider formation energies of BaCuChF and related elemental, binary, and ternary impurities. Formation energies per formula unit of these compounds $\Delta H_F^M$ are summarized in Table I. As expected, the sulfides have more negative formation energy and therefore are more stable than the selenides and the tellurides. Formation energies of BaO, BaCh, and $Cu_2O$ are within 7 % of published experimental results, but for $BaF_2$ and CuF theoretical results are 30% less negative compared to experiment.[30,31]

Second, we consider the full BaCuChF stability regions and choose the sub-regions consistent with experimental synthesis conditions. The stability regions for all three BaCuChF materials are qualitatively similar but the absolute sizes of these regions decrease in the S-Se-Te series. The stability region of BaCuSF shown in Fig. 2a lies close to Cu-rich conditions ((111) plane on the S-Ba-F coordinate system) and is limited to a 3D polyhedron by formation of Cu, BaS, $BaF_2$, CuF, $Cu_2S$ and $BaCu_2S_2$ impurities. Within the BaCuSF stability polyhedron, where the chemical potential of Ba can vary from -7.0 to 0.0 eV, we consider the $\Delta\mu_{Ba} = 0.25\Delta H_F^{BaCuSF}$ = -2.15 eV plane, because Ba is the heaviest atom and its absolute content is the least likely to vary under any experimental conditions (Fig.2b). Our choice is consistent with experimental results, because Fig. 2b predicts the possibility of BaS and $BaF_2$ impurities, and these are the only impurities occasionally observed using XRD in our samples.[3,6]

Third, we choose the chemical potential values for the defect calculations. The BaCuSF stability region within the $\Delta\mu_{Ba} = 0.25\Delta H_F^{BaCuSF}$ plane has three special points. Point *A* is the Cu-rich/Ch-rich/F-poor condition, where BaCuSF is in equilibrium with Cu and BaS; point *B* is the Cu-poor/Ch-rich/F-rich condition, where BaCuSF is in equilibrium with $BaF_2$ and BaS; point *C* is the Cu-rich/Ch-poor/F-rich condition, where BaCuChF is in equilibrium with Cu and $BaF_2$.

We chose to present Cu-rich/Ch-rich/F-poor conditions (point "*A*") in this work, because Ch-rich/F-poor conditions represent typical BaCuChF experimental synthesis conditions that usually involve $H_2Ch$ gas or extra Ch.[3,6], but no additional source of F. The Cu-rich conditions give the lower limit of the free carrier concentration in BaCuChF, since the Cu vacancy was found to be the lowest-energy acceptor-like defect. To estimate the possibility of *p*-type doping with oxygen on the fluorine site, oxygen-related defects were calculated for F-poor conditions (point "*A*"), assuming maximally O-rich conditions defined by equilibrium with the BaO impurity phase. The values of the chemical potentials $\mu_a^0$ and $\Delta\mu_a$ used in the defect calculations are summarized in Table II for *a* = Ba, Cu, Ch, F and O.

B. Native point defects

We calculated formation enthalpies of all vacancies and interstitials of all the atoms on different sites, and cationic and anionic substitutional point defects. Local relaxation of the crystal structure was included in all defect calculations and found to be small except for the $F_{Ch}$ substitutional defect, which we discuss in detail in the context of defect complexes in section III.F. Formation enthalpies $\Delta H_{D,q}$ of the lowest-energy defects for $E_F = 0$ in Eq. 1 are summarized in Table III. The lowest energy acceptor-like defect is $V_{Cu}^{-1}$, and the lowest energy donor-like defect is $V_{Ch}^{+2}$. A recent DFT study of LaOCuSe found that the formation energy of the neutral Cu vacancy is slightly negative, indicating that in LaOCuSe Cu vacancies might by the dominant acceptor like defect, in agreement with our result.[10] Large formation enthalpies of $F_i$ and all the other point defects can be understood considering Eq. 1. The formation enthalpies of the neutral defects depend on the energy needed to distort the bonds (1st term) and on the chemical potential of the atoms (2nd term). Defects have large formation enthalpies if at least one of these terms is large. The 1$^{st}$ term is large if large atoms are added/removed ($V_{Ba}$, $Ba_I$) or if the ionic radius mismatch is large ($Ba_{Cu}$, $Cu_{Ba}$, $F_{Ch}$, $Ch_F$). The 2$^{nd}$ term is large if reactive elements (small $\mu_a^0$, Table I) are removed from the cell ($V_F$, $V_{Ba}$, $Ch_F$, $Cu_{Ba}$, $O_F$) or if benign elements (large $\mu_a^0$, Table I) are added to the cell ($Cu_{Ba}$, $S_F$, $Cu_I$, $S_I$, $O_{Ch}$, $O_I$). All the defects but $V_{Cu}$, $V_{Ch}$, and $F_i$ satisfy at least one of these conditions and therefore have large formation enthalpies. $F_i$ has a large formation enthalpy (Table III) because of the assumed F-poor synthesis conditions (Table II). Under F-rich synthesis conditions $F_I$ has a lower theoretical formation enthalpy, but it remains always above the formation enthalpy of $V_{Cu}$. Therefore, under any synthesis conditions $V_{Cu}^{-1}$ is the most likely source of unintentional *p*-type conductivity in BaCuChF. Unintentional copper vacancies are also known to cause *p*-type conductivity in LaCuOSe,[10] $CuAlO_2$,[17,18] $CuSr_2O_2$,[19] and $Cu_2O$.[32]

Formation enthalpies $\Delta H_{D,q}$ of charged defects depend on the Fermi energy $E_F$ (3$^{rd}$ term in Eq. 1), as shown in Fig. 3 for $V_{Cu}$ and $V_{Ch}$. The formation enthalpy of $V_{Cu}^{-1}$ decreases and the formation enthalpy of $V_{Ch}^{+2}$ increases with increasing Fermi energy, as expected from the sign of the charge of these defects. The transition level $V_{Cu}^{-1/0}$ is below the valence band maximum $E_V$, so most of the $V_{Cu}$ will be charged for any $E_F$ in the gap. In contrast, $V_{Ch}^{0/+2}$, $V_{Ch}^{0/+1}$ and $V_{Ch}^{+1/+2}$ transition levels are deep in the gap. With increasing $E_F$, the doubly-charged $V_{Ch}^{+2}$ state changes to the neutral $V_{Ch}^0$ state without accessing the $V_{Ch}^{+1}$ state, so $V_{Ch}$ is a "negative *U*" center, similar to $V_{Ch}$ in chalcopyrites and $V_O$ in ZnO.[33] Recent studies of $V_O$ in ZnO based on the GW method[34] and of $V_{Cu}$ in $Cu_2O$ using hybrid functionals[32] and using DFT/GGA+U[35] found significant corrections to the transition levels calculated with DFT/GGA. We have not considered approximations that go beyond GGA in this work.

In the S-Se-Te series, the formation enthalpy of the acceptor-like $V_{Cu}^{-1}$ decreases and the formation enthalpy of the donor-like $V_{Ch}^{+2}$ increases (Table III and in Fig. 3). This leads to an enhanced *p*-type character of BaCuTeF compared to BaCuSF and BaCuSeF. Similar S-Se-Te trends are also observed for the other acceptor-like and donor-like defects, respectively. Considering Eq. 1, we attribute these trends to the increasing formation energy of BaCuChF (Table I), and to the increasing energy of its VBM ($E_V$). The VBM in the BaCuChF S-Se-Te series increases due to the increasing energy of the Ch *p*-states and due to increased Ch *p* - Cu *d* repulsion,[36] similar to Cu(InGa)Ch$_2$.[37] The magnitude of these changes in BaCuChF and Cu(InGa)Ch$_2$ should be similar because of their similar valence band character.[26]

Recently, we found that BaCuTeF has a 0.5 eV range of forbidden transitions above the 2.3 eV electronic gap and a 0.2 eV band filling shift.[26] Therefore, the BaCuTeF optical band gap is about 3.0 eV which is comparable to that of wider-bandgap BaCuSF and BaCuSeF. Enhancement of optical transparency due to suppressed probability of the optical transitions in BaCuTeF makes this material similar to CuInO$_2$ (Ref. 38) and In$_2$O$_3$.[39] This character of the near-bandgap region makes BaCuTeF, CuInO$_2$, and In$_2$O$_3$ particularly suitable for optoelectronic applications that require transparency to visible light.

C. Fermi levels

Experimental Fermi levels on the surface (XPS), in the bulk (Hall effect),[3] and theoretical equilibrium Fermi levels (DFT) are close to the valence band maximum and decrease in the S-Se-Te series (Fig. 3 and Table IV). The difference between the experimental surface and bulk Fermi levels indicates that the BaCuChF bands bend down towards the surfaces, which may contribute to scattering of free holes on grain boundaries. Theoretical equilibrium Fermi levels at the 700 °C synthesis temperature that is typically used for the experiments[3] are close to the intersection of the $V_{Cu}^{-1}$ and $V_S^{+2}$ lines, (Fig. 3) which indicates that all three BaCuChF materials are compensated semiconductors, similar to other wide-bandgap semiconductors.[40]

Below we explain how the compensating defects may potentially cause the difficulty of doping BaCuChF and how they cause poor performance of BaCuChF transparent field-effect transistors (TFETs).[40] In the case of *n*-type doping attempts or if a positive electric field is applied to a TFET gate, $E_F$ increases, the formation enthalpy $\Delta H_{D,q}$ of $V_{Cu}^{-1}$ decreases (Eq. 1 and Fig. 3), leading to an increase in the concentration of $V_{Cu}^{-1}$ and additional free holes appear to maintain charge balance, which forces $E_F$ to return to the intersection point at which the charges of $V_{Cu}^{-1}$ and $V_{Ch}^{+2}$ compensate each other. In the case of *p*-type doping attempts, or if a negative electric field is applied to a TFET gate, $E_F$ decreases, free holes disappear to compensate the formation of new $V_{Ch}^{+2}$, and $E_F$ returns to the intersection point. Modulation of the Fermi level in BaCuChF by doping or by electric field is limited by lower and upper "pinning" Fermi levels at which the formation enthalpies of $V_{Ch}^{+2}$ and $V_{Cu}^{-1}$ respectively become negative (Fig. 3 and Table IV). The pinning range (energy difference between the two pinning Fermi levels) in BaCuSF is only 0.2 eV and it further decreases in the S-Se-Te series, which makes it impossible to appreciably modulate $E_F$ in BaCuChF by an applied electric field or by doping.

The Fermi level in BaCuChF can be changed by the formation of an interface with another chalcogenide semiconductor. Deposition of ZnTe on a BaCuSeF surface free of carbon and oxygen contaminants raises the position of the Fermi level to the acceptor pinning Fermi level and leads to formation of copper vacancies and diffusion of resulting copper atoms into the ZnTe interfacial layer.[42] This effect is similar to that of CdS/CuInSe$_2$ interfaces,[43,44] but it was not observed for the BaCuSeF/ZnPc interfaces prepared in a similar way due to trapping of transferred electrons in BaCuSeF surface states.[3]

The non-equilibrium character of the thin film growth gives a limited opportunity to control the Fermi levels in BaCuChF thin films. Cu-rich thin films with a thickness of 400 nm were prepared using an alternating pulsed laser deposition (PLD) technique.[25] Each $BaCu_{1+x}ChF$ sample consisted of 10 periods of alternating BaCuChF and Cu layers that interdiffused during the growth at 400 °C. The $BaCu_{1+x}ChF$ films exhibited phase pure XRD patterns with no Cu diffraction peaks. As the amount of excess Cu in $BaCu_{1+x}SF$ increased to $x$=0.05 (determined by EPMA), both resistivity $\rho$ and Seebeck coefficient $S$ increased by about a factor of 5, which may be attributed to partial filling of copper vacancies due to increase of the copper chemical potential. However, for larger amount of excess Cu, $\rho$ and $S$ decreased, which was attributed to formation of Cu-related impurity bands as manifested by moderate sub-gap optical absorption. Optical properties of these samples are discussed in the next section. We conclude that it is difficult to control the concentration of free holes in BaCuChF using either equilibrium or non-equilibrium synthesis approaches, or by electric field gating.

D. Optoelectronic properties

First, we consider optical absorption spectra of the specially prepared Cu-rich BaCuChF thin films and S-poor BaCuSF thin films, as verified using EPMA. Fig. 4a and 4b show the difference in the optical absorption spectra of $BaCu_{1+x}SF$ and BaCuSF thin films and the difference in the optical absorption spectra of $BaCu_{1+x}TeF$ and BaCuTeF thin films, respectively. The experimental setup is schematically shown in the inset of Fig. 4a, and the proposed model for the $BaCu_{1+x}ChF$ sub-gap optical absorption is shown in the inset of Fig. 4b. Sub-gap absorption peaks labeled $E_X$, $E_Y$, and $E_Z$ are attributed to optical transitions to the conduction band minimum from Cu-related defect bands close to the valence band maximum. The exact origin of these impurity bands remains unclear.

The absorption spectrum of a defective BaCuS$_{1-x}$F thin film deposited at 500 °C (Fig. 4c) consists of three broad bands labeled $E_A$, $E_B$, and $E_O$. As shown in the inset of Fig. 4c, these absorption bands are attributed to direct excitation from the valence band to donor-like $V_{Ch}^{+1}$, $V_{Ch}^{+2}$ and $O_{Ch}^{+1}$ defect states, respectively. As the BaCuS$_{1+x}$F sample ages in air, intensities of all the absorption peaks decrease, but the absorption strength of the $E_O$ peak decreases more slowly than the $E_A$ and $E_B$ peaks. We attribute the increase of the $E_O/E_A$ and the $E_O/E_B$ intensity ratios to filling of sulfur vacancies with atmospheric oxygen. This interpretation is consistent with the proposed absorption model.

Second, we discuss the results of photoluminescence (PL) measurements on normal BaCuChF thin films without intentional Cu excess and without intentional S deficiency (Fig. 5). The PL measurement setup is schematically depicted in the inset of Fig. 5b. The emission bands $E_A$, $E_B$, and $E_O$ are within 0.1eV of the measured absorption peaks (Fig. 4c) and therefore are likely to originate from recombination through donor-like $V_{Ch}^{+1}$, $V_{Ch}^{+2}$ and $O_{Ch}^{+1}$ defects. The proposed PL mechanism is shown in the inset of Fig. 5c. Electrons photo-excited to the conduction band by UV light are trapped by positively-charged donor-like defects and recombine with free holes in the valence band, emitting a photon. This PL model is consistent with the experiments, since the $E_O/E_A$ and $E_O/E_B$ relative band intensities are larger in the BaCuSF powder aged for 5 years in air (Fig. 5c) than in the freshly prepared BaCuSF thin film (Fig. 5a).

Additional support for the recombination to the valence band maximum is provided by the low-energy shoulder of the $E_A$ peak in the BaCuSeF (Fig. 5b). The energy difference between this shoulder and the main $E_A$ peak is 0.1 eV, which is consistent with the spin-orbit splitting of the valence band in BaCuSeF.[26]

Third, we compare the experimental results for absorption and PL to the DFT defect calculation results. Experimental and DFT results agree that the presence of sulfur vacancies causes tensile strain, measured in S-poor BaCuSF thin films reported previously[25] and calculated using DFT supercells with $V_{Ch}$ reported in this work. Also, persistent photoconductivity (PPC) was observed in BaCuSF (Fig. 5a, inset) and BaCuSeF thin films, which may be a signature of $V_{Ch}$, similar to $V_{Ch}$ in chalcopyrites and $V_O$ in oxides.[33] Spectral positions of the main absorption and PL peaks in BaCuSF and BaCuSeF are within 0.4 eV of the calculated $V_{Ch}^{+1/0}$, $V_{Ch}^{+2/+1}$ and $O_{Ch}^{+1/0}$ transition levels. A large discrepancy is expected, because the transition levels were determined from the result of two fully relaxed defect calculations with different defect atomic configurations, which is not the case for the fast optical processes of photon absorption or emission.[33]

BaCuTeF thin films and powders did not show PL and PPC because of the absence of Te character in the conduction band minimum (CBM).[26] In the PL process, a photoexcited electron at the CBM of BaCuTeF is spatially separated from $V_{Te}$, and therefore its trapping probability is much lower than in BaCuSF and BaCuSeF materials that have Ch-character in the CBM. It is likely that recombination of electron-hole pairs in BaCuTeF is nonradiative.

E. Oxygen

Oxygen is a likely contaminant and a possible dopant in BaCuChF materials.[3,6] Oxidation of unprotected BaCuChF surfaces was observed using optical absorption (Fig. 4), PL (Fig. 5), and XPS measurements.[3] According to photoelectron spectroscopy measurements, surface oxidation decreases the BaCuChF work function[45] which inhibits *ex-situ* processing of this material in thin film solar cells.[2,46]

The combination of oxidized grain boundaries and anisotropic crystal structure causes a 27 fold decrease in mobility of polycrystalline BaCuTeF thin films compared to the *c*-axis oriented epitaxial samples.[28] Since the ratio of average to in-plane effective mass is 3,[26] the additional factor of 9 must stem from the difference in the carrier mean free path in polycrystalline and epitaxial samples caused by the oxidation of the grain boundaries. In contrast to BaCuChF, grain boundary scattering does not affect the mobility of polycrystalline films of isostructural BiCuOCh materials.[47] The benign character of BiCuOSe grain boundaries may be caused by a smaller amount of Bi-O in the grain boundaries and better transport through them, as compared to wider-bandgap Ba-O and La-O in BaCuChF and LaCuOCh grain boundaries, respectively. We conclude that mobility of free holes in polycrystalline thin films of chalcogenide materials composed of cations that are benign to oxidation (Bi, Ni, Cu, *etc.*) and that have naturally *p*-type conductive oxides (BiO, NiO, $Cu_2O$, *etc*) is less likely to be affected by grain boundary scattering compared to Ba- and La-based materials. A good example of this principle is provided by the benign nature of grain boundaries in $Cu(InGa)Se_2$ chalcopyrites.[48] This conclusion provides chemical guidance for the design of polycrystalline *p*-type transparent chalcogenides with mobility of free holes suitable for optoelectronic applications.

Despite easy surface oxidation in BaCuChF, *p*-type doping with oxygen on the fluorine site in related SrCuChF materials is experimentally challenging.[49] According to the present DFT calculations on BaCuChF, the formation enthalpy of $O_F^{-1}$ is smaller than for $O_{Ch}^{+1}$, but larger than for $V_{Cu}^{-1}$ (Table V) even under the favorable O-rich/F-poor/Cu-rich conditions (Table II). The equilibrium concentration of $O_F^{-1}$ ($10^{15}$-$10^{17}$ cm$^{-3}$) is much smaller compared to that of $V_{Cu}^{-1}$ ($10^{21}$ cm$^{-3}$). Under O-rich/Ch-poor conditions, $O_{Ch}^{+1}$ becomes more abundant that $O_F^{-1}$, which explains the oxidation of BaCuChF surfaces upon exposure to an oxidizing environment such as air. We conclude that doping with oxygen on the fluorine site in BaCuChF is not likely and that oxygen impurities do not affect the concentration of free holes in bulk BaCuChF, but decrease their mobility due to grain boundary oxide scattering.

F. Defect complexes

BaCuChF materials have several possible defect complexes that are more energetically favorable than the isolated point defects (Table VI). Formation of neutral $[V_{Ch}^{2+}+2V_{Cu}^{-1}]^0$ defect complexes is likely, because equilibrium concentrations of $V_{Cu}^{-1}$ and $V_{Ch}^{+2}$ are large ($10^{20}$ - $10^{21}$ cm$^{-3}$), and because both defects are located in the same layer of the BaCuChF crystal structure (Fig. 1). A similar neutral defect compensation mechanism is also observed in other multicomponent semiconductors, for example ($2V_{Cu}^{-1}+In_{Cu}^{+2}$) in CuGaSe$_2$,[19] and ($Cu_{Zn}^{1-}+Zn_{Cu}^{1+}$) in Cu$_2$ZnSnS$_4$.[20,21]

Defect pairs in BaCuChF can also be formed by relaxation of the $F_{Ch}$ substitutional defects towards the Ba plane. The lowest energy configuration of atoms due to this relaxation resembles a positively charged $[V_{Ch}^{+2}+F_I^{-1}]^{+1}$ defect complex. This defect complex can trap an electron, in which case electronic states appear ~1 eV below the conduction band minimum. Large relaxation and deep gap state are signatures of the DX centers in binary[50] and ternary[51] chalcogenides, so $[V_{Ch}^{+2}+F_I^{-1}]^{+1}$ is likely to act as DX center in BaCuChF materials. $[V_{Ch}^{+2}+F_I^{-1}]^{+1}$ defect pairs along with $V_{Cu}^{-1}$ can also form neutral $[V_{Ch}^{+2}+F_I^{-1}+V_{Cu}^{-1}]^0$ defect complexes. Both $[V_{Ch}^{+2}+F_I^{-1}]^{+1}$ and $[V_{Ch}^{+2}+F_I^{-1}+V_{Cu}^{-1}]^0$ defect complexes have negative formation energy $\delta H_{DC,q}$ compared to formation energy of the isolated point defects (Table VI). The energy gain to form all the considered defect complexes decreases in the S-Se-Te series, which is consistent with the fact that compensating defects are more likely to form in wide-bandgap semiconductors than in low-bandgap materials.[41] Additional energy gain from forming defect complexes in BaCuChF may come from their ordering, as happens in Cu(InGa)Se$_2$ chalcopyrites.[15] The possibility of formation of ordered defect complexes and an intrinsic DX center in BaCuChF materials warrants a more detailed study beyond the scope of this publication.

## VI. SUMMARY AND CONCLUSIONS

We have reported results of DFT calculations of native point defects, defect complexes and oxygen impurities in BaCuChF semiconductors. Unintentional *p*-type conductivity and Fermi level pinning in BaCuChF is explained by a defect model comprised of $V_{Cu}^{-1}$ and $V_{Ch}^{+2}$ defects. The proposed model is consistent with experimental results of electrical transport, optical and photoemission spectroscopy measurements. Results of this work lead to the conclusion that BaCuChF materials are suitable for transparent *p*-type contact applications, but not suitable for fabrication of transparent field-effect transistors with *p*-type channels.

## ACKNOWLEDGMENTS

This work was supported by the National Science Foundation of USA under Grant No. DMR-0804916. We thank Henri Jansen and David Roundy for useful discussions on DFT, Gregory Rorrer, Debra Gail and Oksana Ostroverkhova for their assistance with the PL measurements and interpretation, Evan F. DeBlander and Morgan Emerson for their help with the thin film characterization, Douglas A. Keszler and Cheol-Hee Park for BaCuChF powder

TABLE I Formation energies per formula unit of BaCuChF and related materials

| $\Delta H_F^M$ (eV) | Ch=S | Ch=Se | Ch=Te |
|---|---|---|---|
| BaCuChF | -8.60 | -8.28 | -7.98 |
| BaCu$_2$Ch$_2$ | -5.20 | -4.54 | -3.95 |
| BaCh | -4.45 | -4.17 | -3.75 |
| Cu$_2$Ch | -0.45 | -0.03 | +0.26 |
| Fluorides: | BaF$_2$:-7.97; CuF:-1.59 | | |
| Oxides: | BaO:-5.75; Cu$_2$O:-1.32 | | |

TABLE II Chemical potentials used for the defect calculations in BaCuChF

| $\mu$ (eV) | Ch=S | Ch=Se | Ch=Te |
|---|---|---|---|
| $\Delta\mu_{Ba}$ | -2.15 | -2.07 | -2.00 |
| $\Delta\mu_{Cu}$ | -0.00 | -0.00 | -0.00 |
| $\Delta\mu_{Ch}$ | -2.34 | -2.10 | -1.74 |
| $\Delta\mu_F$ | -4.11 | -4.11 | -4.25 |
| $\Delta\mu_O$ | -3.60 | -3.68 | -3.76 |
| $\mu^0_{Ba}$=-1.92; $\mu^0_{Cu}$=-3.64 | | | |
| $\mu^0_F$=-1.80; $\mu^0_O$=-4.42 | | | |
| $\mu^0_S$=-3.91; $\mu^0_{Se}$=-3.51; $\mu^0_{Te}$=-3.05 | | | |

TABLE III Formation enthalpies of the most likely native point defects in BaCuChF at $E_F$=0.

| $\Delta H_{D,q}$ (eV) | Ch=S | Ch=Se | Ch=Te |
|---|---|---|---|
| $V_{Cu}^0$ | 0.34 | 0.27 | 0.18 |
| $V_{Cu}^{-1}$ | 0.22 | 0.13 | -0.03 |
| $V_{Ch}^0$ | 5.78 | 4.40 | 3.46 |
| $V_{Ch}^{+1}$ | 3.11 | 2.61 | 2.16 |
| $V_{Ch}^{+2}$ | -0.01 | 0.09 | 0.20 |
| $F_i^0$ | 1.67 | 1.62 | 1.56 |
| $F_i^{-1}$ | 1.52 | 1.50 | 1.42 |

TABLE IV Fermi levels in BaCuChF

| $E_F$ (eV) | Ch=S | Ch=Se | Ch=Te |
|---|---|---|---|
| DFT (700°C) | 0.10 | 0.03 | -0.05 |
| DFT (300°C) | 0.07 | 0.05 | 0.04 |
| pin. acceptors | 0.22 | 0.13 | -0.03 |
| pin. donors | 0.01 | -0.05 | -0.10 |
| XPS (300°C) | 0.30 | 0.20 | 0.10 |
| Hall (300°C) | 0.03 | -0.05 | -0.23 |

TABLE V Formation enthalpies of point oxygen impurities in BaCuChF at $E_F=0$.

| $\Delta H_D$ (eV) | Ch=S | Ch=Se | Ch=Te |
|---|---|---|---|
| $O_F^0$ | 1.55 | 1.46 | 1.10 |
| $O_F^{-1}$ | 1.23 | 1.21 | 0.87 |
| $O_{Ch}^0$ | 3.93 | 3.06 | 2.96 |
| $O_{Ch}^{-1}$ | 2.66 | 2.52 | 2.26 |
| $O_{Ch}^{+1}$ | 1.81 | 1.32 | 2.14 |

TABLE VI Energy costs to form defect complexes in BaCuChF at $E_F = 0$.

| $\delta H_{DC,q}$ (eV) | Ch=S | Ch=Se | Ch=Te |
|---|---|---|---|
| $[V_{Ch}+2V_{Cu}]^0$ | -1.33 | -1.21 | -0.91 |
| $[V_{Ch}+F_I]^{+1}$ | -2.75 | -2.57 | -1.30 |
| $[V_{Ch}+V_{Cu}+F_I]^0$ | -2.75 | -2.57 | -1.30 |

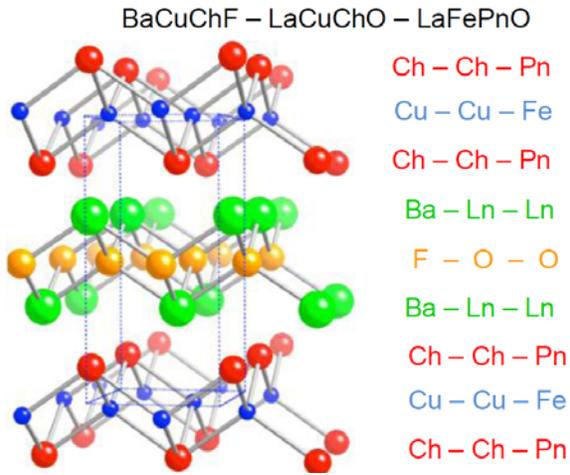

FIG. 1 Crystal structure of BaCuChF (Ch = S, Se, Te), LnCuOCh (Ln = La, Pr, Nd) and LaFeOPn (Pn = P, As, Sb).

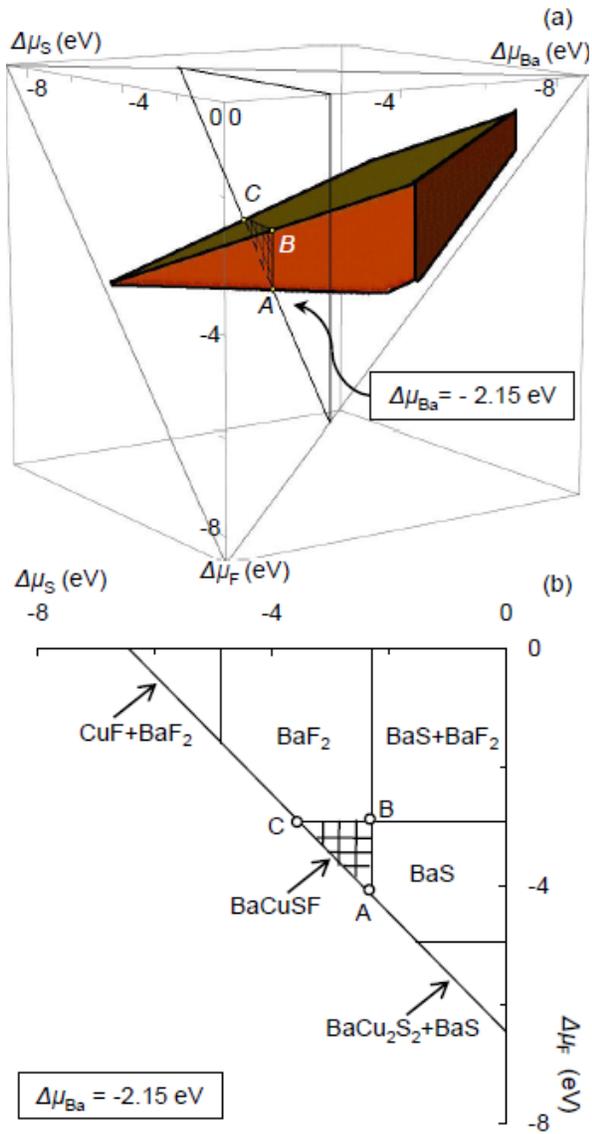

FIG. 2 (a) BaCuSF stability polyhedron (b) Crossection of the stability polyhedron by $\Delta\mu_{Ba}=0.25\Delta H^{BaCuSF}$ plane.

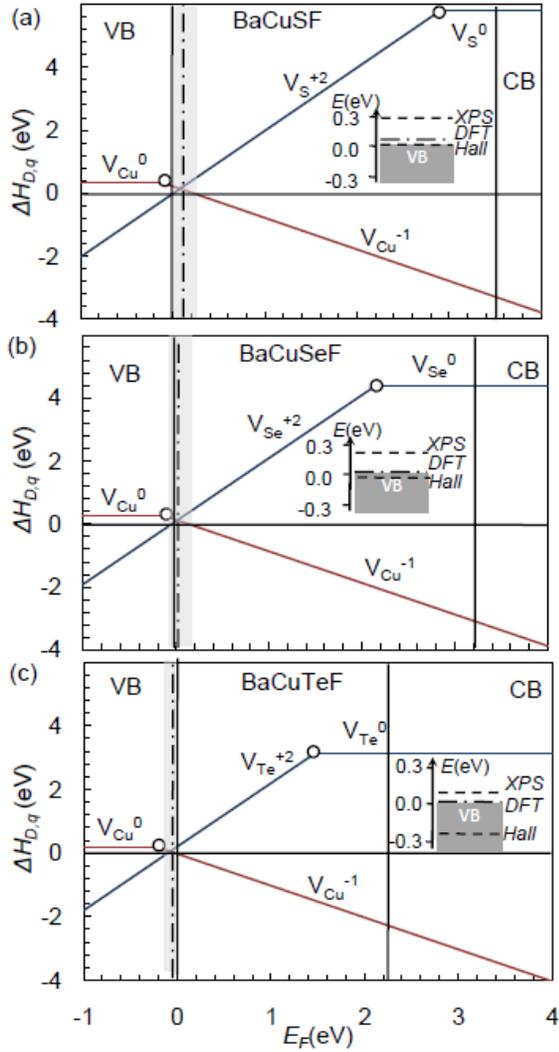

FIG. 3 Formation enthalpies of the lowest charge state of $V_{Cu}$ and $V_{Ch}$ as a function of Fermi energy for (a) BaCuSF, (b) BaCuSeF and (c) BaCuTeF. The empty circles at the kinks of the lines correspond to defect transition levels. The Fermi energy pinning range is shadowed. The vertical dash-dotted lines are the equilibrium Fermi energies at 700 °C. Insets: Experimental Fermi energies on the surface (XPS), in the bulk (Hall), and theoretical equilibrium Fermi energies (DFT) at 300 °C.

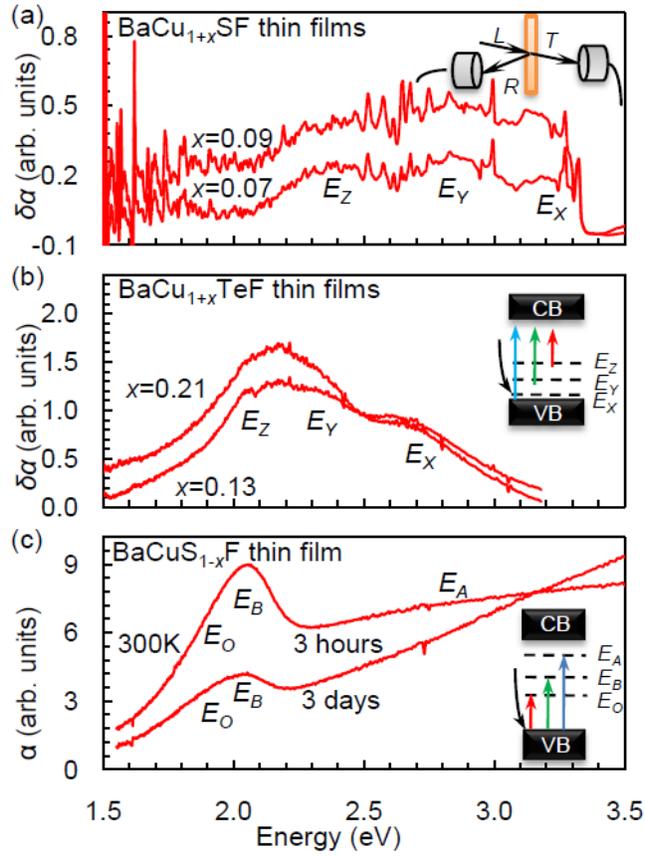

FIG. 4 Differential optical absorption spectra of $BaCu_{1+x}ChF$ thin films for (a) Ch=S, (b) Ch=Te. (c) Optical absorption spectra of a $BaCuS_{1-x}F$ thin film measured 3 hours and 3 days after removal from vacuum. Insets: (a) absorption measurement setup, (b) proposed $BaCu_{1+x}ChF$ absorption model, (d) proposed $BaCuS_{1-x}F$ absorption model.

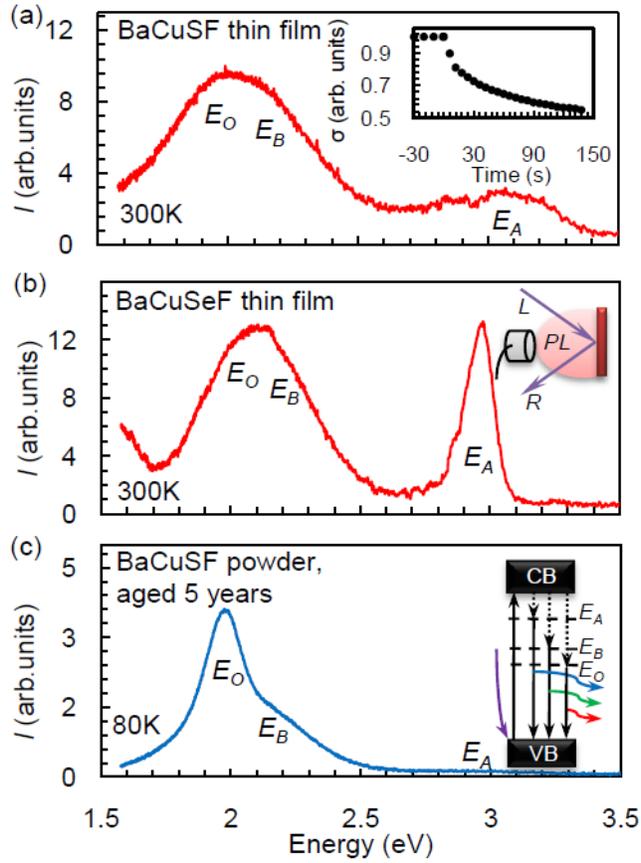

FIG. 5 Photoluminescence spectra of (a) BaCuSeF thin film, (b) BaCuSF thin film and (c) BaCuSF powder aged for 5 years. Insets: (a) normalized conductivity of BaCuSF thin film after light exposure, (b) PL measurement setup, (c) proposed PL model.